\documentclass{pasj00}
\draft

\usepackage{longtable}
\usepackage{color}

\begin{document}
\SetRunningHead{Hiroko Watanabe}{Observations of Umbral Dots}
%\Received{}%{yyyy/mm/dd}
%\Accepted{}%{yyyy/mm/dd}
%\Published{}%{yyyy/mm/dd}

\title{Observations of Umbral Dots and their Physical Models}

%%% begin:list of authors
% Do NOT capitalize all letters in "textsc".
\author{Hiroko \textsc{Watanabe}} %
%  \thanks{Example: Present Address is xxxxxxxxxx}}
\affil{Unit of Synergetic Studies for Space, 
	Kyoto University, Kitashirakawa-Oiwake-cho, Sakyo-ku,  
      Kyoto 606-8502, JAPAN}
\email{watanabe@kwasan.kyoto-u.ac.jp}

%% `\KeyWords{}' always has to be placed before `\maketitle'.
\KeyWords{sunspots -- Sun: magnetic fields -- convection -- Sun: photosphere} %Do NOT move this preamble from here!

\maketitle

\begin{abstract}
The Hinode satellite opens a new era to the sunspots research, because of its high 
spatial resolution and temporal stability. 
Fine scale structures in sunspots, called umbral dots (UDs), 
have become one of the hottest topics in terms of the close observation  
of the magnetoconvection.  

In this paper, a brief review of observed properties of UDs is given based 
on the recent literature. 
UDs born in the periphery of the umbra exhibit inward migration, 
and their speeds are positively correlated with the magnetic field inclination.
Longer-lasting UDs are tend to be larger and brighter, while 
lifetimes of UDs show no relation with their background magnetic field strength.  
UDs tend to disappear or stop its proper motion by colliding with 
the locally strong field region. 
The spatial distribution of UD is not uniform over an umbra 
but is preferably located at boundaries of cellular patterns.  
From our 2-dimensional correlation analysis, we 
measured the characteristic width of the cell boundaries ($\approx$0.5{\arcsec}) 
and the size of the cells ($\approx$6{\arcsec}).

Then we performed a simplified analysis to get statistics how the UD 
distribution is random or clustered using the Hinode blue continuum images.  
We find a hint that the UDs become less dense and more clustered for later 
phase sunspots.  These results may be related to the evolutional change of 
the subsurface structure of a sunspot.     

Based on these observational results, 
we will discuss their physical models 
by means of numerical simulations of magnetoconvection.
\end{abstract}

\section{Introduction}

The observation and analysis of sunspots have been performed 
passionately from the age of Galileo. 
Although we have an accumulation of more than 400 year's 
observational data, the sunspot remains to be one of the biggest unsolved 
problem in the solar physics:  
We still do not reach a consensus about 
the subsurface structure of a sunspot, i.e., whether the magnetic field is clustering or monolithic
\citep{1972SoPh...26...52G, 1979ApJ...230..905P, 2003A&ARv..11..153S}. 
How their strong magnetic field are born and 
what determines its lifetime?\citep{2010ApJ...720..233C, 2011ApJ...740...15R}
Understanding the sunspot is highly important for the astrophysics, 
because violent solar activities in the strong magnetic field of sunspot are driven 
by a common mechanism with many cosmic eruptive events 
(accretion jets, stellar flares, auroras, ...). 

\begin{figure}
 \begin{center}
  \includegraphics[width=7cm]{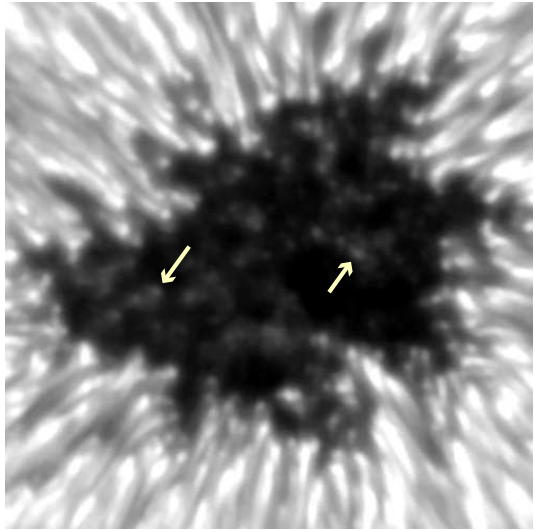} 
 \end{center}
\caption{Blue continuum image of NOAA 10944 observed by the 
Hinode SOT on 2007 March 2.  The FOV is about 17{arcsec}$\times$17{arcsec}. 
The arrows indicate two prominent UDs in the umbra.}\label{fig:UD}
\end{figure}

We find a new path for solving these problems in the research of fine scale bright 
points called umbral dots (UDs, Figure\,\ref{fig:UD}, reviews are found in 
\cite{2011LRSP....8....4B}). 
UDs are transient brightenings observed in sunspot
umbrae and pores, with typical scales of 300~km diameter and 10~min 
lifetime \citep{1997A&A...328..682S, 1997A&A...328..689S, 2007PASJ...59S.585K, 
2008A&A...492..233R, 2012ApJ...752..109L}. 
The magnetic properties around UDs are studied by many authors 
\citep{2008ApJ...678L.157R, 2009ApJ...694.1080S, 2012ApJ...757...49W}, 
and they reach a consensus that UDs exhibit local reduction of field strength at deep layer. 
Upflow inside UDs and confined down-flowing regions outside of them 
are observed in recent high-resolution observations
(Ortiz et al. 2010, Watanabe et al. 2012, \cite{2013A&A...554A..53R}). 

A UD is considered to be a natural consequence of the interaction 
between the convection and the magnetic field based on the monolithic sunspot model 
\citep{2006ApJ...641L..73S, 2010A&A...510A..12B}.
One of the most sophisticated computer simulation of 
magnetoconvection performed by 
\citet{2012ApJ...750...62R} succeeded in reproducing the overall 
structure of the sunspot.   
The convective motions inside the umbra push out the boundary of 
the magnetic field lines inside the convective cell, creating a region of 
strongly reduced field strength, and forming a cusp or canopy field 
configuration.
As UDs are strongly linked with the subsurface 
through an interaction with the deep convective layer, 
they have a potential to offer us information of 
the unreachable subsurface structure and its dynamics. 
  
The paper is organized as follows.  The review of recent observational 
analyses, and our new analysis on the UD distribution are given in Section~\ref{sec:review_analysis}.   In
Section~\ref{sec:model}, we discuss how these observational results are 
consistently interpreted by means of numerical studies of magnetoconvection.
Finally, based on the analysis and discussion, we conclude how the 
UD analysis can be one of the most important topics in the solar physics in Section~\ref{sec:conclusion}.

\section{Review of Recent Analyses}\label{sec:review_analysis} 

The observation of fine scale structure like UDs needs 
high spatial resolution. 
The Hinode Solar Optical Telescope (SOT) has a main mirror of 
50-cm aperture and achieves its diffraction limit of 0.2-0.3\arcsec\ 
ALWAYS because of its seeing-free environment \citep{2008SoPh..249..167T, 
2008SoPh..249..197S}. 
This temporal stability is the best advantage for reliable analysis on the 
structure's temporal evolution. 
On the other hand, the Swedish Solar Telescope (SST, \cite{2003SPIE.4853..341S}) 
has an effective aperture of 1-m, twice as large as that of the Hinode.  
Plus they possess a spectro-polarimetric 
imaging system called CRISP, which enables a scanning sequences 
of magnetic sensitive lines rapidly only in a few minutes intervals. 

The analyses below utilizes the most optimum observations ever 
for their individual purposes. 

\subsection{Inward migration}\label{sec:inward_migration} 

UDs show an apparent motion of so-called inward migration toward 
the center of the umbra. 
An important question we have to answer is 
that why the UDs migrate always inward to the umbra center
, and why they are better seen in the umbral periphery. 
\citet{2009ApJ...702.1048W} analyzed the apparent motion speed 
of more than 2000 UDs and compared them to their background 
umbral field inclination. 
The field inclination is vertical at the center of the umbra, and 
becomes more inclined at the periphery. 
A positive correlation between the field inclination 
and speed of UDs are found (Figure\,\ref{fig:inclination_speed}). 
Least-square linear fitting for the samples with lifetime longer than 
100~s gives us the following relations:  \\
$V$  = 0.014 $\times$ $i$ (field inclination [degree]) + 0.070 \\
and\\
$V$ = $-0.49$ $\times$ $B$ (field strength [kG]) + 1.5\\
where $V$ is the apparent motion speed in unit of km s$^{-1}$.

\begin{figure}
 \begin{center}
  \includegraphics[width=8cm]{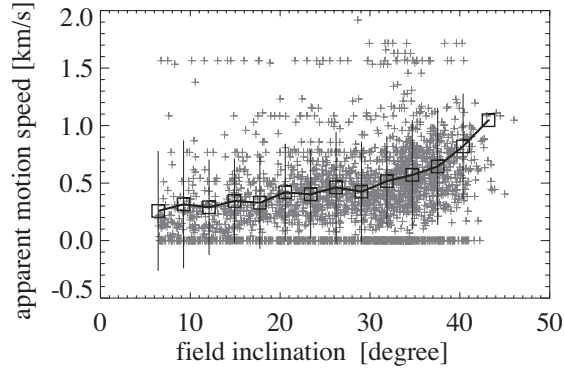} 
 \end{center}
\caption{Scatter plots of field inclination versus the apparent motion 
speed of 2268 UD samples.  
The average bins of 3$^{\circ}$ is shown 
with square symbols, and vertical solid line denote the standard error deviation 
errors. Modified from Figure\,4 in \citet{2009ApJ...702.1048W}.}\label{fig:inclination_speed}
\end{figure}

\subsection{``Parameter survey'' of magnetoconvective manifestation}\label{sec:magnetoconvection}  

As UDs are manifestation of magnetoconvection and they occur in 
various environmental magnetic field background,  
the observational characteristics of UDs can be used as ``parameter survey'' of magnetoconvection. 
Here we show some important plots showing how UD's lifetime or 
radius depends on its environmental umbral field strength (Figure\,\ref{fig:inclination_speed}). 

\begin{figure}
 \begin{center}
  \includegraphics[width=0.9\textwidth]{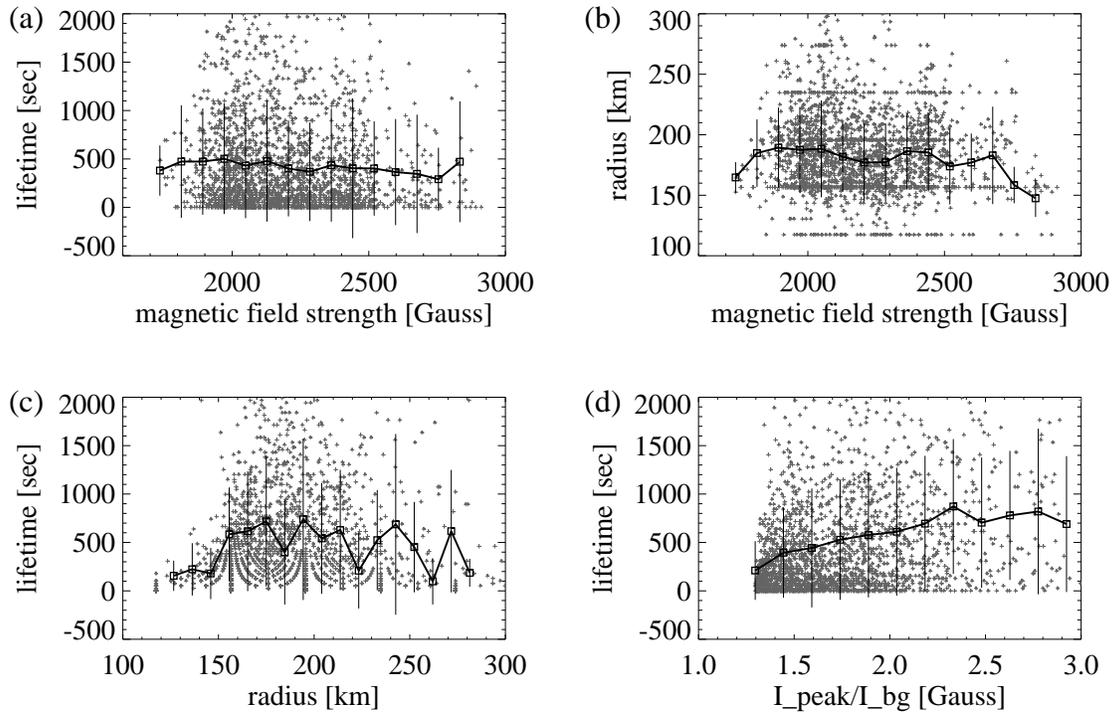} 
 \end{center}
\caption{Scatter plots of (a) magnetic field strength versus lifetime (b)  
magnetic field strength versus radius (c) radius versus lifetime (d) peak intensity at UD normalized 
by the dark background intensity versus lifetime.  There are 2268 UD samples. 
The averages for a proper binning size are shown 
with square symbols, and vertical solid line denote the standard error deviation 
errors. Extended from Figure\,4 in \citet{2009ApJ...702.1048W}.}\label{fig:inclination_speed}
\end{figure}

The lifetime of UDs scatter a lot and the average  
is almost constant regardless of the field strength and field inclination (not shown). 
The radius or the area of UDs, on the other hand, exhibits a weak hint of 
getting smaller in very strong ($>$2600 Gauss) fields. 
This is readily understood by the inhibition of expansion of 
buoyant plasma in the presence of the strong magnetic field pressure 
\citep{2013A&A...551A..92T}. 

\subsection{Evolution tracking}\label{sec:evolution_insitu}

\begin{figure}
 \begin{center}
  \includegraphics[width=9cm]{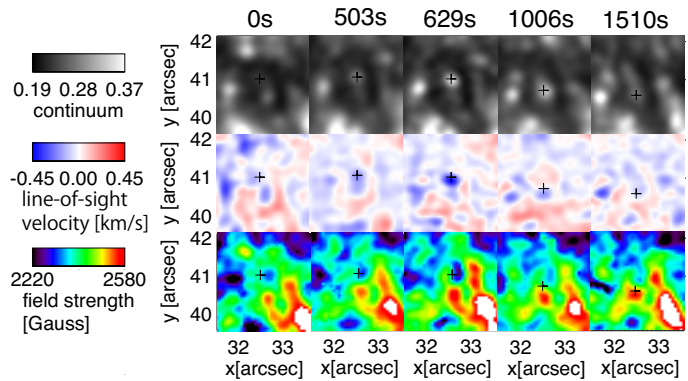} 
 \end{center}
\caption{The temporal evolution of one central UD in continuum (top), line-of-sight velocity 
(middle), and magnetic field strength (bottom).  The FOV is about 3{\arcsec}$\times$3.5{\arcsec}. 
The elapsed time from the emergence is 
shown at the top of the images.  The black plus symbol indicate the UD position detected by 
the peak of the continuum.  This UD disappears 1573\,s after its emergence. 
Modified from Figure\,13 in \citet{2012ApJ...757...49W}.}\label{fig:UD_detail}
\end{figure}

A detailed statistical work on the temporal evolution of UDs was 
performed by \citet{2012ApJ...757...49W}. 
From the spectropolarimetric observation taken at SST/CRISP, 
maps of line-of-sight velocity and vector magnetic field around UDs are obtained. 
One example for a UD appeared in the central region of the umbra is shown in Figure\,\ref{fig:UD_detail}. 
By seeing many UD movies, we found the following characteristics: 
\begin{itemize}
\item The peak of continuum intensity and line-of-sight velocity coincides in time. 
\item The locations of UD appearance exhibit weakening of field strength compared 
with their surroundings in the growing phase of the UD. 
On the other hand, in the diminishing phase, UDs tend to collide into the pre-existing 
locally strong field regions. 
\item For the evolution of migrating UDs (e.g., Figure\,17 in \cite{2012ApJ...757...49W}), 
they are preferentially located at the boundary of weak and strong field
regions, and the boundary evolves with UD's migration as if the leading edge of the UD 
is always blocked by strong field ``walls''. 
\end{itemize}

\citet{1995ApJ...447L.133S} reported that the proper motion of 
peripheral or grain-origin UDs are influenced by the dark umbral cores: 
UDs slowed down and disappeared at the borders of dark umbral cores. 
Our third result is consistent with their result if we consider the dark umbral 
cores as the locally strong field regions.

\subsection{Analysis on the theoretical UDs}\label{sec:Bharti} 

A sophisticated work combining the observational data analysis with the 
simulated UDs was performed by \citet{2010A&A...510A..12B}. 
They first performed a 3-dimensional MHD simulation in a box filled by 
a fixed vertical field flux which corresponds to a mean field strength of 2500~G. 
Afterwards they applied a typical data analysis method on the simulated UDs. 
In this way, a direct compari?son with simulation and observation becomes 
possible. 

The bottom panels (c) and (d) of Figure\,\ref{fig:inclination_speed} are 
comparable to Fig.10 in \citet{2010A&A...510A..12B}. 
The radius or the area of smaller UD ($<$180\,km) has a positive correlation between 
lifetime (c.f., the mean radius of our 2268 UDs is 183\,km.), showing consistency
with \citet{2010A&A...510A..12B}.  
However above 180\,km radius, the correlation is missing.   
The lifetime become longer for brighter UDs, as is also consistent with 
Bharti's results. 
Similar trends of size-lifetime and brighteness-lifetime relations are suggested by 
\citet{2009A&A...504..575S}, while \citet{2008A&A...492..233R} 
do not find any systematical trend.

We can not compare Bharti's result with the correlations in 
panels (a) and (b) in Figure\,\ref{fig:inclination_speed},  
because the simulation pose a fixed vertical flux. 
However by M. Sch{\"u}ssler \& A. V{\"o}gler (2006, private communication), 
the lifetime decreases as field strength gets larger if the amount of 
the fixed vertical flux is altered, which contradicts with the observation.
On the other hand, the Stanford university group performed a numerical 
MHD simulations on solar magnetoconvection and found the opposite 
lifetime dependency, i.e., lifetime increases as field strength gets larger 
(Irina. N. Kitiashvili, private communication). 
Both the observation and simulation need more verifications to conclude this discussion. 

\subsection{Subsurface diagnosis}\label{sec:subsurface}  

\subsubsection{Cellar pattern}
 
\begin{figure}
 \begin{center}
  \includegraphics[width=8cm]{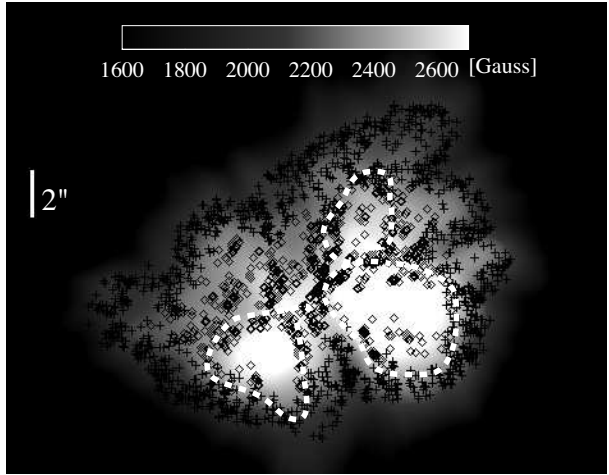} 
 \end{center}
\caption{The emergence positions of 2268 UD samples 
overlaid on the background image of magnetic field strength.  
The thick dotted lines contours over some of the 
cellular patterns. 
Modified from Figure\,6 in \citet{2009ApJ...702.1048W}. 
(color-version figure is available in the on-line material.)}\label{fig:distribution}
\end{figure}

\citet{2009ApJ...702.1048W}  showed the map of the emergence positions 
of more than 2000 UDs (Figure\,\ref{fig:distribution}).   
From Figure\,\ref{fig:distribution}, we can recognize some cellular 
patterns in their distribution overlaid by thick dashed lines. 
These cellular patterns remind us of the cluster-type 
magnetic configuration \citep{1979ApJ...230..905P}, which 
the fields are separated into several bundles.  
Possibly they are the remnant of pores getting together  
in the developing phase of this sunspot. 
\citet{2002A&A...388.1048T} and \citet{1997A&A...328..689S} also 
concluded that the spatial distribution of UDs is not random but avoids 
dark umbral cores.
On the other hand, \citet{2008A&A...492..233R} made a similar map of 
UD's distribution and it showed a uniform distribution. 
At this point we do not reach to a consensus yet.  

We applied two-point correlation function method in order to quantify 
how the distribution is different from completely random one 
as a function of mutual distances between UDs.  
This method is also used for galaxies distribution observed by 
the Sloan Digital Sky Survey \citep{1980lssu.book.....P}. 
While the galaxies are in the 3-dimensional distribution, 
our UD distribution is in 2-dimensional plane. So   
the two-point correlation function can be reduced: 
\begin{equation}
P(x_{1},x_{2})dx_{1} dx_{2}=\bar{n} [1 + \xi(r)] dx_{1} dx_{2}
\end{equation}
$P(x_{1},x_{2})dx_{1} dx_{2}$ is the probability that the both finite 
areas around the two point ($x_{1}$ and $x_{2}$, $r=|x_{2}-x_{1}|$) 
 contains samples. 
 $\bar{n}$ is the average number density of UDs, and 
 $\xi(r)$ is called the two-point correlation function. 
If the distribution is completely random $\xi(r)$ is always zero. 
$\xi(r)$ can take negative values if the correlation is weaker. 
Or if we rewrite in the term of $r$ and the number of samples ($dN$)
within the annular zone between $r$ and $r+dr$ around one sample, then 
\begin{equation}
dN = 2\pi r \bar{n} dr [1+\xi(r)]
 \end{equation}
It is important to accumulate $dN$ for as many samples as possible and 
calculate their average ${\overline{dN}}$ to get reliable statistics.  

\begin{figure}
 \begin{center}
  \includegraphics[width=9cm]{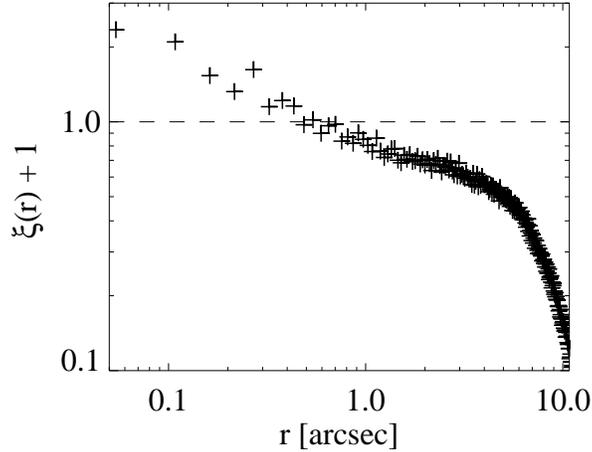} 
 \end{center}
\caption{Plots of two point correlation function ($\xi(r)+1$) versus distance ($r$) for 
the distribution of UDs in Figure\,\ref{fig:distribution} in log-log scale.}\label{fig:twopointcorrelation}
\end{figure}

We applied this two-point correlation function calculation to 
the distribution of Figure\,\ref{fig:distribution}, and the result 
is shown in Figure\,\ref{fig:twopointcorrelation}. 
We can get two hints from this plot. 
First, $\xi(r)$ is positive within distance smaller than 0.5\arcsec. 
This is about 3 times longer than the typical radius of UDs, and may correspond 
to the element size of the clustering of UD occurrence. 
Or in other words, 0.5\arcsec\ corresponds to the width of 
of the cell boundaries or walls.  
Second, there is a broken branch of the log-log linear relation 
at around 6\arcsec. 
It tells that there are a change in the pattern above the size of 6\arcsec. 
Indeed the approximate diameter of the largest cellular pattern have the 
the diameter of $\approx$6\arcsec. 
$\xi(r)$ rapidly decreases as it goes beyond 10\arcsec, and this is because 
it gets close to the boundary of the sunspot periphery whose diameter is about 16\arcsec.

\subsubsection{Statistical analysis of UD's distribution}

Here we perform another simplified analysis to describe the statistical 
characteristic of UD's spatial distribution.  
We chose 16 sample sunspots arbitrary from 2013 June to August 
taken by the Hinode SOT blue continuum filtergrams. 
For each sample, we analyze only one snapshot image. 
All images have spatial pixel scale of 0.109\,\arcsec.
The UD positions in each image are measured by eye within regions of 
continuum intensities darker than 0.4 $\times$ $I_{\textrm{qs}}$ (quiet-sun intensity) 
of background images (defined by smoothing the original image by 10~pixel $\times$ 
10~pixel width).  
The area of the region, number of detected UDs, number density 
of UDs are listed in table~\ref{tbl:tbl2}.  
Then we estimate the uniformity of the UD distribution.  
The concept is similar to the nearest neighbor method comparing to the Poisson distribution 
described in \citet{1954Ecol...35..445C}, but our uniformity ratio (U) is more appropriate for 
our spatially discrete samples  
\begin{equation}
U=\frac{(\textrm{mean distance between nearest neighbors})}{2\sqrt{\frac{{\textrm{(area)}}}{\pi{\textrm{(number of sample)}}}}}
\end{equation}
The concept of $U$ comes from the following context; 
Suppose the whole area is completely packed with small circles of 
the area divided by the number of sample.  The diameter of this small circles is 
expressed as the denominator of  the equation (3). If each sample is located 
at the center of these circles, this is the mostly scattered and uniform distribution, which is $U$=1.
If $U$ is smaller than 1, the population is clustered.    
Typically the error of the detected number of UDs is about 5\%, 
then the typical error of the ratio $U$ is about 2.5\%.  

Also in table~\ref{tbl:tbl2}, we list up the phase of sunspot evolution classified roughly into four phases 
(0: just emerged, 1: evolving, 2: mature, 3: disintegrating) by tracking 
its temporal evolution using the helioviewer (http://www.helioviewer.org).  
Rough estimation of magnetic field strength is obtained by the  
daily sunspot drawings at the 150-Foot Solar Tower at Mt. Wilson Observatory 
(ftp://howard.astro.ucla.edu/pub/obs/drawings, \cite{2009ASSL..361.....V}) in a 
unit one hundred gauss.  

\begin{longtable}{ccccccc}
\caption{Results of sunspot distribution analysis (during 2013 June-August)}\\
\hline
\hline
 time in UT & ($x$,$y$) &  sunspot  & num. of & num. density  & $U$ & mag. field \\
 & [arcsec]  &  phase$^{*A}$ & UDs & [arcsec$^{-2}$] & ratio & [$\times$100\,G]$^{*B}$\\
\hline
\endhead
\hline
\multicolumn{7}{c}{
\begin{minipage}{20cm}\vspace*{3mm} 
{\footnotesize *A: 0: just emerged, 1: evolving, 2: mature, 3: disintegrating,} \\
{\footnotesize *B:  V: negative polarity, R: positive polarity} \end{minipage} 
} 
\endfoot
\small{Jun-14 10:38:06} & (540, -210)  & 1 & 27 & 0.25 & 0.49 & V23 \\
\small{Jun-14 10:38:06} & (590,-200) & 1 & 27 & 0.48 & 0.50 & R23  \\
\small{Jun-15 09:23:20} & (680,-220) & 2 & 25 & 0.36 & 0.50 & V20  \\
\small{Jun-15 09:23:20} & (740,-210) & 2 & 27 & 0.55 & 0.47 & R21  \\
\small{Jul-05 02:51:27} & (-530,-260) & 2 & 64 & 0.19 & 0.45 & R24  \\
\small{Jul-06 02:35:51} & (-330,-260) & 3 & 64 & 0.21 & 0.47 & R24  \\
\small{Jul-25 02:43:03} & (750,250) & 2 & 12 & 0.30 & 0.54 & V23  \\
\small{Jul-25 05:09:34} & (270,-300) & 1 & 16 & 0.63 & 0.55 & R19  \\
\small{Jul-25 05:09:34} & (290,-250) & 1 & 38 & 0.63 & 0.58 & V22  \\
\small{Jul-25 05:09:34} & (330,-220) & 1 & 11 & 0.59 & 0.56 & R19  \\
\small{Jul-25 05:09:34} & (380,-245) & 1 & 29 & 0.65 & 0.64 & R20  \\
\small{Jul-27 02:23:04} & (640,-240) & 2 & 55 & 0.82 & 0.53 & V20  \\
\small{Aug-01 20:53:34} & (80,-360) & 0 & 12 & 0.56 & 0.45 & R20  \\
\small{Aug-17 18:42:01} & (470,-220) & 2 & 119 & 0.50 & 0.52 & R24  \\
\small{Aug-20 01:29:50} & (820,-180) & 3 & 34 & 0.31 & 0.42 & R24  \\
\small{Aug-21 06:29:04} & (-31,-250) & 2 & 71 & 0.51 & 0.54 & R24  
\label{tbl:tbl2}
\end{longtable}

\begin{figure}
 \begin{center}
  \includegraphics[width=0.9\textwidth]{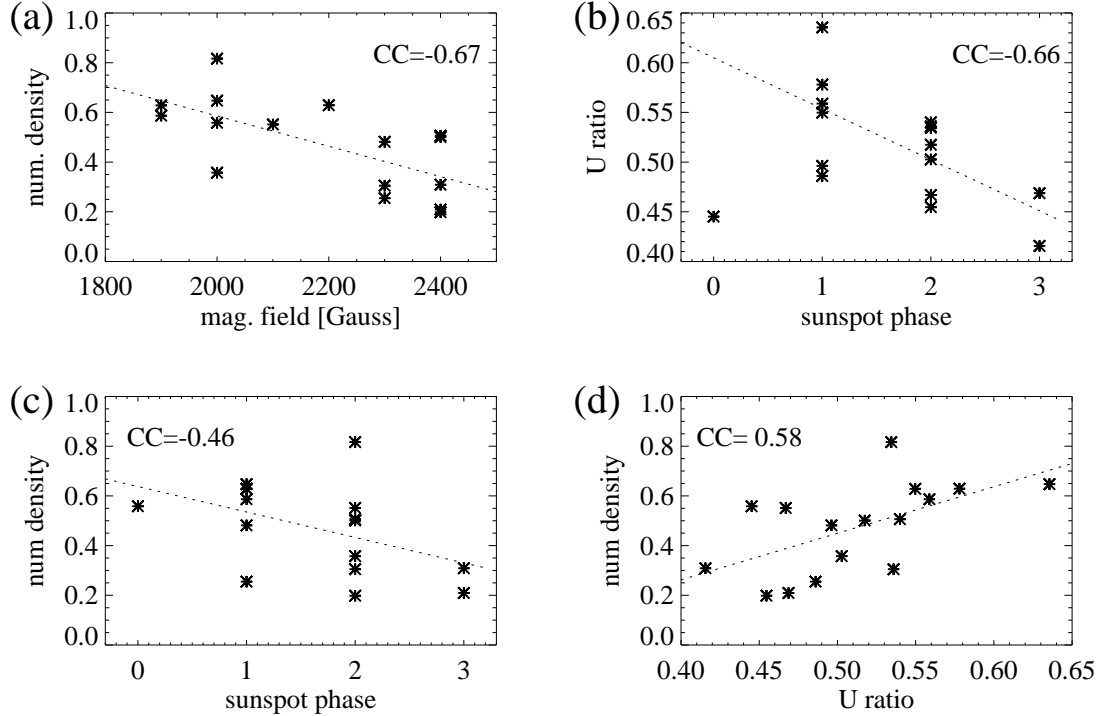} 
 \end{center}
\caption{Scatter plots of (a) the magnetic field strength versus number density (b)  
the phase of sunspot evolution versus $U$ ratio (c) the phase of sunspot evolution versus 
number density (d) $U$ ratio versus number density.  The dotted lines are linear 
fitting, and the number CC shows the correlation coefficient between the two 
variables. For the panel (b), both the fitting and CC are calculated excluding one 
data point at phase=0.}\label{fig:scatterplot}
\end{figure}

Some scatter relations are shown in Figure\,\ref{fig:scatterplot}. 
Although with large scatters, we found some hints that: 
\begin{itemize}
\item Number density of UDs decreases as the magnetic field gets stronger. 
\item The distribution of UDs become more clustered (smaller $U$) for later phase sunspot 
than for earlier phase sunspot (except for one sample at phase = 0).  
\item Later phase sunspots have smaller number density. 
\item Distribution becomes more uniform (large $U$) for larger number density sunspot. 
\end{itemize} 

These statistical correlations suggest us a working view on 
the evolution of sunspot internal structures.  When a sunspot 
is evolving, the umbral magnetic field is not uniform as the umbra 
is formed by aggregation of many pores with a variety of field strength.  
The spatial scale of inhomogeneity is small corresponding to the 
reminiscent of original components. 
At this phase, magnetic field is relatively weak, UDs are 
numerous, and distribute nearly uniformly.  As the spot evolves, 
the small scale inhomogeneity of field strength gradually smoothed 
and strengthened, leading to the formation of large scale dark umbral 
cores where UDs are suppressed to appear.  The degree of clustering will be large. 
Confirmation of this working view remains to the future more detailed analysis.

\section{Model of UDs}\label{sec:model}  

The numerical simulation of UDs was successfully performed 
by \citet{2006ApJ...641L..73S} in the framework of the magnetoconvection 
based on the monolithic sunspot model. 
Magnetoconvection can explain various manifestations 
in sunspots (UDs, light bridges, penumbral filaments) in a common way 
\citep{2006A&A...447..343S, 2006A&A...460..605S}.  
Here we discuss the consistency between the model and the 
observational results. 

We see many samples showing smooth transition from penumbral grains to UDs.  
The magnetoconvection model explains the different appearance of 
these structures by the change of convection cell shape in a large field inclination 
\citep{2009ApJ...700L.178K, 2007ApJ...669.1390H, 2011SoPh..268..283K}. 
In a large inclination situation, the convection cells become
elongated while in a smaller inclination they are circular.   

The inward migration is interpreted as the successive appearances of 
hot plasma plumes \citep{2008ApJ...677L.149S}.  
When hot gas plume ascends from below, it bends the magnetic field 
line down (more inclined) at the surface locally, 
reducing the magnetic pressure there, and triggering a new appearance of hot gas. 
As the field is making in a fan-shape, their direction is always 
toward the center of the umbra. 
Accordingly it is readily understandable that there is positive correlation 
between the field inclination and the speed of inward migration, because 
this bending process does not occur in a vertical field situation.  

The trend that longer-lasting UDs are brighter and larger is natural, 
if we suppose stronger and larger convective upflows for those UDs. 
However there is a discrepancy between observation and theory in 
the lifetime versus field strength relation (section\,\ref{sec:Bharti}). 
There are too large scatters in UD's lifetime distribution in our observation. 
More precise definition of lifetime or larger statistics may 
give some relations as a future work. 

The trend of disappearance/stopping at locally strong field region 
may correspond to the stronger suppression of convection there. 
However it is not easy to imagine the existence of strong field 
``wall'' (boundary of weak and strong field) 
at the leading edge of migrating UD, which also migrate 
along with the motion of UD (section~\ref{sec:evolution_insitu}).
Here we remind of the flux tube model \citep{1998A&A...337..897S}
which appeared for explaining penumbral Evershed flow by siphon flow mechanism.  
This model do not explain UDs, but it covers the mechanism of the penumbral grains.   
In the flux tube model, single flux tube located at the magnetopause is heated 
and get buoyant, and the footpoint of buoyant flux tube goes 
inward to the umbra. This migration of footpoint may correspond 
to the inward migration of the leading edge of penumbral grains.  
The strong field ``wall'' can be imagined as the compression of 
preexisting umbral field by the the buoyant flux tubes.  

\section{Conclusion}\label{sec:conclusion} 

A UD is a unique cosmic phenomenon for understanding magnetoconvection 
with both spatially and temporarily resolved observations. 
Such studies that compare with the numerical simulation and 
the observed properties is possible owing to this rich data set 
\citep{2010A&A...510A..12B, 2012ApJ...745..163K}. 
There are other magnetoconvection manifestation in the 
universe, such as the accretion disk and the low temperature stars, 
although those are very far and so tracking the temporal evolution 
of their magnetoconvective cells is almost impossible.  
The observation of UD, in a sense, can be called as the ``laboratory'' 
of the magnetoconvection. 

In this paper, we suppose the usability of a UD as a subsurface diagnosis. 
The most popular tool for a subsurface diagnosis is the helioseismology 
\citep{2010ARA&A..48..289G, 2010SoPh..267....1M}. 
However recently, \citet{2013A&A...558A.130S} looked into the ability of 
helioseismology to probe the subsurface of sunspots, and they found that 
the subsurface magnetic field configuration below 2~Mm is not sensitive 
to the method of the helioseismology.  
We test the performance of two-point correlation function for the 
distribution of UD's occurrence positions, and found a weak hint of 
cellular patterns with 6\arcsec\ diameter and 0.5\arcsec\ annular width. 
These patterns may corresponds to the flux bundles well below the sunspot 
as the cluster model predict. 
Although the well-known simulations of UDs (e.g., \cite{2006ApJ...641L..73S}) are based on the 
monolithic sunspot model, we could not rule out the cluster sunspot model 
because UDs occur in shallow layers below umbra surface in their their simulations 
($<$1\,Mm), and their connection to deeper layer structure is still under investigation.
We also introduce a simple parameter $U$ for describing the uniformity of the 
UD's distribution.  
Our results tell that the distribution of UDs is more uniform in the earlier phase, 
and become more clustered in the later phase. 
It may represent that the global subsurface structure of a sunspot changes as time, 
because UDs occur in the interaction between the deep convective layer and 
magnetic field of a sunspot.  
The statistical analysis of UD's distribution have a large potential to draw out 
information of the subsurface structure of a sunspot, 
and thus is worth to be extended further in the near future.

%%%%%%%%%%%%%%%%%%%%%%%%%%%%%%%%%%%%%%%

\bigskip

H.~Watanabe wants to thank here Dr.~Reizaburo Kitai, Dr.~Kiyoshi Ichimoto, 
and Dr.~Luis R. Bellot Rubio for their collaborative and supportive work. 
Our work was supported by the Grant-in-Aid for JSPS Fellows, and by the JSPS Core-to-Core
Program 22001. 
Hinode is a Japanese mission developed and launched by ISAS/JAXA, with 
NAOJ as domestic partner and NASA and STFC (UK) as international partners. 
It is operated by these agencies in co-operation with ESA and NSC (Norway).
This research has made extensive use of NASA's Astrophysical Data System.

%%%
% See the manual for the detail.
%%%

\end{document}